\documentclass[aps,prd,preprint,superscriptaddress,tightenlines,nofootinbib]{revtex4}

\usepackage{graphicx}
\usepackage{dcolumn}
\usepackage{bm}

\begin{document}

\preprint{CLNS 03/1824}       
\preprint{CLEO 03-08}         

\title{Search for Baryons in the Radiative Penguin Decay
$b \rightarrow s \gamma $}



\author{K.~W.~Edwards}
\affiliation{Carleton University, Ottawa, Ontario, Canada K1S 5B6 \\
and the Institute of Particle Physics, Canada}
\author{D.~Besson}
\affiliation{University of Kansas, Lawrence, Kansas 66045}
\author{S.~Anderson}
\author{V.~V.~Frolov}
\author{D.~T.~Gong}
\author{Y.~Kubota}
\author{S.~Z.~Li}
\author{R.~Poling}
\author{A.~Smith}
\author{C.~J.~Stepaniak}
\author{J.~Urheim}
\affiliation{University of Minnesota, Minneapolis, Minnesota 55455}
% ...
\collaboration{CLEO Collaboration} 
\noaffiliation


\date{May 2, 2003}

\begin{abstract} 
We have searched for the baryon-containing radiative penguin decays $B^-
\rightarrow \Lambda \overline{p} \gamma$ and $B^- \rightarrow \Sigma^0
\overline{p} \gamma$, using a sample of $9.7 \times 10^6$ $B\overline{B}$
events collected at the $\Upsilon(4S)$ with the CLEO detector.  We find
no evidence for either, and set 90\% confidence level upper limits of 

$$[{\cal B} (B^- \rightarrow \Lambda \overline p \gamma) + 0.3
{\cal B} (B^- \rightarrow \Sigma^0 \overline p \gamma)]
_{E_\gamma > 2.0 \rm{~GeV}}  < 3.3 \times 10^{-6}\ ,$$

$$[{\cal B} (B^- \rightarrow \Sigma^0 \overline p \gamma) + 0.4
{\cal B} (B^- \rightarrow \Lambda \overline p \gamma)]
_{E_\gamma > 2.0 \rm{~GeV}}  < 6.4 \times 10^{-6}\ .$$

\noindent From the latter, we estimate

$${\cal B} (B \rightarrow X_s \gamma, X_s ~containing ~baryons)
_{E_\gamma > 2.0 \rm{~GeV}}  < 3.8 \times 10^{-5}\ .$$

\noindent This limit implies upper limits on corrections to CLEO's recent
measurement of branching fraction, mean photon energy, and variance in
photon energy from $b \rightarrow s \gamma$ that are less than half the
combined statistical and systematic errors quoted on these quantities.

\end{abstract}

\pacs{13.20.He, 13.40.Hq, 13.60.Rj}
\maketitle


The branching fraction for the radiative penguin decay $b
\rightarrow s \gamma$ has been shown to place significant 
restrictions on physics
beyond the Standard Model (SM) \cite{Hurth}. The
photon energy spectrum, in contrast, is insensitive to beyond-SM
physics \cite{Anatomy}, but provides information on the $b$ quark
mass and
momentum within the $B$ meson, information useful for determining
the CKM matrix elements $\vert V_{ub}\vert$ and $\vert
V_{cb}\vert$. Some measurements of ${\cal B} (b \rightarrow s
\gamma)$ \cite{old,Belle}, including the most precise one to date
\cite{new}, and the best measurement of the photon energy spectrum
\cite{new} include a technique (\lq\lq pseudoreconstruction") that
has reduced sensitivity to those $B \rightarrow X_s \gamma$ decays
with baryons in the final state. It is therefore important to
determine what fraction of $B \rightarrow X_s \gamma$ decays lead
to baryons, or to place an upper limit on that fraction.

Calculations of the mass distribution of the $s \bar q_{spectator}$
system that hadronizes into $X_s$
\cite{Ali-Greub,Anatomy} show 1/3 of the spectrum above 2.05 GeV/$c^2$,
the  threshold for $\Lambda \overline p$ (the lightest
baryon-containing final state), so a sizeable rate for $b
\rightarrow s \gamma$ with baryons would not be unexpected. Of
the spectrum above $\Lambda \overline p$ threshold,
2/3 is below 2.5 GeV/$c^2$, so one expects the baryon-containing final
states to be dominated by $\Lambda \overline N$ and
$\Sigma \overline N$.
Thus, measurements of the branching fractions for $B \rightarrow \Lambda
\overline{N} \gamma$ and $B \rightarrow \Sigma \overline{N} \gamma$ would
help estimate a correction to the $b \rightarrow s \gamma$ branching
fraction and photon energy spectrum.  In addition, the decay $B
\rightarrow \Lambda \overline{N} \gamma$ provides a method for
determining the helicity of the photon in $b \rightarrow s \gamma$.  For
$\Lambda \overline{N}$ systems near threshold (s-wave), or for $\Lambda$
and $\overline{N}$ near back-to-back to the photon (thus orbital angular
momentum perpendicular to the photon), the $\Lambda$ and $\gamma$ have
the same helicity.  A measurement of the $\Lambda$ helicity, via its
decay angle distribution, gives a measurement of the $\gamma$ helicity. 
We have therefore conducted searches for $B^-
\rightarrow \Lambda \overline p \gamma$
and $B^- \rightarrow \Sigma^0 \bar p \gamma$ and their charge 
conjugates.

 The data used for this analysis were taken with the
CLEO detector \cite{CLEO-detector} at the Cornell
Electron Storage Ring (CESR), a symmetric $e^+e^-$ collider, 
and consist of 9.1 fb$^{-1}$ on the $\Upsilon (4S)$ resonance
($9.7 \times 10^6 ~B \overline B$ events) and 4.4 fb$^{-1}$
60 MeV below the resonance. We select hadronic events that contain a
$\Lambda \rightarrow p \pi^-$, a $\overline p$, and a high energy
photon ($E_{\gamma}^{lab} >$1.5 GeV), or contain a $\overline \Lambda$, a
$p$, and
a high energy photon. (Henceforth, charge conjugate modes are
implied.) For the $B^- \rightarrow \Sigma^0 \bar p \gamma$ search
we {\it do not} reconstruct the $\Sigma^0$, but analyze as if the
decay were $B^- \rightarrow \Lambda \bar p \gamma$, not detecting the
soft photon from $\Sigma^0 \rightarrow \Lambda \gamma$.
The high energy photon must lie in the central region of the
calorimeter $(\vert \cos \theta_\gamma \vert < 0.7)$, must not form a
$\pi^0$ or $\eta$ meson with any other photon
in the event, and must have a lateral energy distribution in the
calorimeter 
consistent with that for a photon. The $\Lambda$ requirements
involve significance of displacement of vertex from interaction
point and consistency of $dE/dx$ and time of flight of the decay proton
candidate with expectation.  They result in a $\Lambda$ candidate sample
that is 90\% pure.  The antiproton candidate must pass $dE/dx$ and
time of flight requirements and must not form a $\overline \Lambda$
with any $\pi^+$ candidate in the event.

We compute the standard $B$ reconstruction variables $M_{\rm cand}
\equiv \sqrt{E^2_{\rm beam} - P^2_{\rm cand}}$ and $\Delta E
\equiv E_{\rm cand} - E_{\rm beam}$, keeping for further analysis
events with $M_{\rm cand} > 5.0$ GeV/$c^2$ and $\vert \Delta E \vert <
0.5$ GeV. $P_{cand}$ and $E_{cand}$ are computed from $\Lambda$,
$\bar p$, and $\gamma$ only, both for
$B^- \rightarrow \Lambda \bar p \gamma$ and
$B^- \rightarrow \Sigma^0 \bar p \gamma$.
With this event selection, there is negligible
background from other $B$ decay processes, but substantial
background from continuum processes: initial state radiation,
photons from decays of $\pi^0$ or $\eta$ that have escaped the
veto, photons from decays of other hadrons. To suppress the
continuum background, we compute twelve event shape variables and
apply loose cuts on three of them.  The twelve variables are then used 
as inputs to a neural net.  The net is trained to distinguish between
signal and continuum background using Monte Carlo samples of each.  
Monte Carlo samples distinct from those used to train the net are used to
determine that cut on neural net output which would give the lowest upper
limit on the branching fraction should the branching fraction actually be
zero, and also that cut which would allow us to see the smallest possible
signal.  Those two cuts differ little, and we use their average.

The event shape variables are calculated in two frames of reference, the
lab frame and the primed frame, the frame of the
system recoiling against the photon.  Variables in the primed frame are
better at rejecting initial state radiation; those in the lab frame are
better at rejecting other continuum events.
The twelve input variables to the neural net are 1) $\vert \cos
\theta_{tt} \vert$, where $\theta_{tt}$ is the angle between the
thrust axis of the candidate $B$ and the thrust axis of the rest
of the event, calculated in the lab frame; 2) $\vert \cos
\theta^\prime_{tt} \vert$, the same, but calculated in the primed frame;
3) the thrust of the candidate $B$; 4) the thrust of the rest of the event;
5) $R_2$, the ratio of second and zeroth Fox-Wolfram \cite{FW} moments,
calculated in the lab frame; 6) $R_2^\prime$, the same, but calculated
in the primed frame; 7) $\vert \cos \theta^\prime \vert$, where
$\theta^\prime$ is the angle between the photon and the thrust
axis of the rest of the event, all calculated in the primed frame;
8) and 9) energies in 20$^\circ$ and 30$^\circ$ cones about the photon
direction (excluding the photon energy); 10) the sum of the magnitudes
of the component of momentum perpendicular to the thrust axis of the
candidate $B$, for particles more than 45$^\circ$ from this axis,
divided by the sum of the magnitudes of momentum for all
particles, in both sums excluding particles from the candidate
$B$, in the lab frame; 11) the same, but evaluated in the primed
frame; and 12) $\cos \theta_B$, where $\theta_B$ is the angle between
the beam direction and the direction of the candidate $B$.

The loose cuts are $R_2 < 0.5$, $R_2^\prime < 0.3$, $\vert \cos
\theta_{tt} \vert < 0.8$.  Having obtained substantial suppression
of background with the loose cuts and the cuts on the net output,
our final selection is from the $2D$ distribution in $M_{\rm cand}
- \Delta E$ space. We define a \lq\lq signal box" $\vert \Delta E
\vert < 84$ MeV, $\vert M_{\rm cand} - M_B \vert < 8$ MeV$/c^2$, which,
based on Monte Carlo simulation, should contain $\sim$90\% of the
$B^- \rightarrow \Lambda \bar p \gamma$
signal events and $(0.75 \pm 0.15)$\% of the background events.
We use the yield of events in the large $M_{\rm cand} -\Delta
E$ region (excluding the signal box), $M_{\rm cand} > 5.0$ GeV/$c^2$,
$\vert \Delta E \vert < 0.5$
GeV, to predict the background in the signal box.  For
$B^- \rightarrow \Sigma^0 \bar p \gamma$, we shift the signal box
by 114 MeV to negative $\Delta E$, compensating for the missing soft
photon from $\Sigma^0 \rightarrow \Lambda \gamma$.  The shifted signal
box should contain $\sim$80\% of the
$B^- \rightarrow \Sigma^0 \bar p \gamma$ signal events.

\begin{figure*}
\begin{center}
\includegraphics*{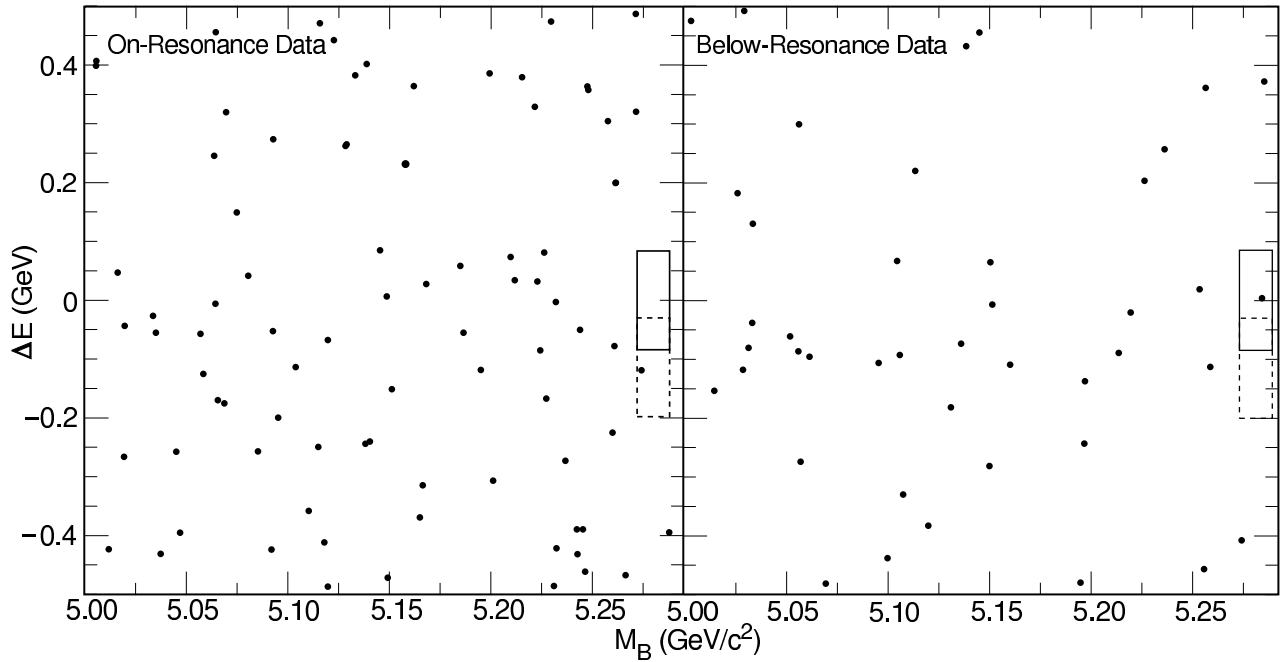}
\caption{\label{fig:onoff}$M_B - \Delta E$ for On and Below resonance
data and photons with $E_{\gamma}^{lab} > 1.5$ GeV.
The solid box shows the tight signal box ($5.272 < M_B < 5.288$ GeV/$c^2$,
$|\Delta E| < 0.084$ GeV) used for determining the $B^- \rightarrow
\Lambda \overline{p} \gamma$ yield.  The dashed box is shifted downward
in $\Delta E$ by 114 MeV and is used for determining the $B^- \rightarrow
\Sigma^0 \overline{p} \gamma$ yield.}
\end{center}
\end{figure*}

The $2D$ distributions in $M_{\rm cand} - \Delta E$ space, On-4S
resonance and Below-resonance, are shown in Fig.~\ref{fig:onoff}.   
There are 84 events On-resonance, and 43 events Below-resonance (with
$\sim$ half the luminosity), leading to a background prediction of 0.6
events On, 0.3 events Below, in either signal box. In the $B^- \rightarrow
\Lambda \overline{p} \gamma$ signal box, we observe zero events On and
one event Below.  In the $B^- \rightarrow \Sigma^0 \bar p \gamma$ signal
box, we observe one event On and zero events Below.
The one On event has a $B$ rest frame photon
energy of 2.18 GeV, estimated by imposing the constraints that the
undetected $\Sigma^0 \rightarrow \Lambda \gamma$ decay photon brings
$\Delta E$ to zero, and combines with the $\Lambda$ to give the
$\Sigma^0$ mass.  Thus, we have no evidence for $B^-
\rightarrow \Lambda \overline p \gamma$, and have a 90\%
confidence level upper limit on its true mean of 2.30 events.  For
$B^- \rightarrow \Sigma^0 \bar p \gamma$, with one event observed and 
a background of 0.6 events expected, we also have no evidence of signal.
We use the pre-Feldman-Cousins PDG procedure\cite{PDGlim} for calculating
upper limits.  We allow for a systematic error in background by
conservatively using only half the expected background in the upper
limit calculation.  That gives a ``conservative 90\% confidence level''
upper limit of 3.64 events.  With the additional requirement that the $B$
rest frame photon energy be greater than 2.0 GeV, the background 
in the large $M_{cand} - \Delta E$ region drops to
27 events On and 15 events Below, with 0.21 background events predicted
for the $B^- \rightarrow \Sigma^0 \bar p \gamma$ signal box.  This leads
to an upper limit of 3.80 events for $E_\gamma >$ 2.0 GeV.  

The upper limit on the branching fraction will be those upper limits
on the number of signal events,  divided by the
number of charged $B$'s, by the $\Lambda \rightarrow p \pi^-$ branching
fraction, and by the detection efficiency.

We assume equal number of charged and neutral $B$'s, noting that a
correction for this assumption can be applied at such time as the
$B^+ B^-$ to $B^0 \overline {B^0}$ ratio in $\Upsilon (4S)$ decays
has been well determined. Thus, we have 9.7 million charged $B$'s.
We assign a $\pm$2\% uncertainty to that number.

We use Monte Carlo simulation to determine the efficiency for detecting $B^-
\rightarrow \Lambda
\overline p \gamma$, and $B^- \rightarrow \Sigma^0 \bar p \gamma$.
The Standard Model predicts left-handed photons and $s$ quarks in $b
\rightarrow s \gamma$, and
thus the $\Lambda$'s in $B^- \rightarrow \Lambda \bar p \gamma$ will
tend to be left-handed.  This tendency decreases the number of pions
from $\Lambda \rightarrow p \pi^-$ that decay against the $\Lambda$
boost direction.  Such pions are soft and more difficult to detect.  Thus, SM
decays will have a detection efficiency {\it higher than} that of
unpolarized $\Lambda$'s.  For an upper limit, we conservatively
assume unpolarized $\Lambda$'s.  For
$B^- \rightarrow \Sigma^0 \bar p \gamma$, we also assume unpolarized
$\Lambda$'s.

\begin{figure}
\begin{center}
\includegraphics*[width=3.35in]{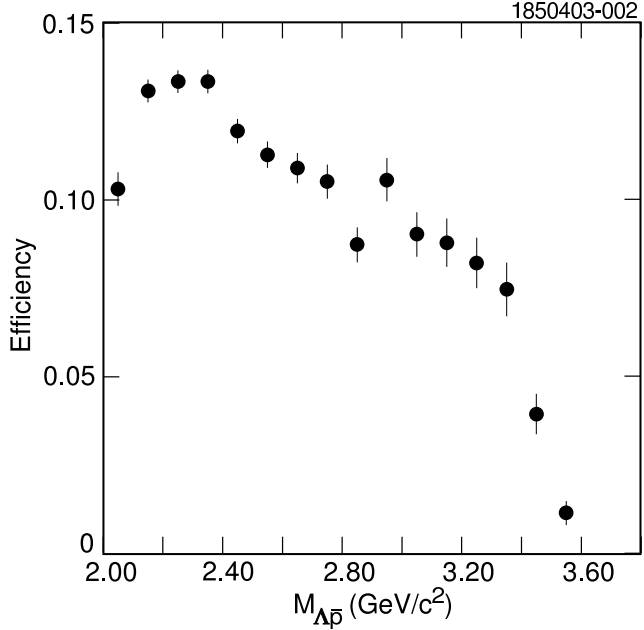}
\caption{\label{fig:effmlp}Efficiency as a function of $M_{\Lambda \bar
p}$, for signal Monte Carlo, with all cuts applied.}
\end{center}
\end{figure}

The efficiency as a function of $\Lambda \bar p$ mass 
is shown in
Fig.~\ref{fig:effmlp}.  The sharp fall-off near 3.5 GeV/$c^2$ is caused by the
photon
energy requirement, $E_{\gamma}^{lab} > 1.5$ GeV.  The gentler decrease
from 2.4 to 3.4 GeV/$c^2$ is caused by the background suppression
requirements.  Similar results are obtained for $\Sigma^0 \overline{p}$.
We assume a $\Lambda \bar p$ mass distribution
($\Sigma^0 \bar p$ mass distribution)
given by the parton-level hadronic mass distribution
\cite{Ali-Greub,Anatomy} times a phase space factor $P/M$, $P$
being the momentum of $\Lambda$ or $\bar p$ ($\Sigma^0$ or $\bar p$)
in the $\Lambda \bar p$ ($\Sigma^0 \bar p$)
rest frame, for that value of $\Lambda \bar p$ ($\Sigma^0 \bar p$)
mass $M$.  We have
also used a weighting $P^3/M$, appropriate for a p-wave system.

As we are primarily interested in decays with a high energy photon,
we compute the efficiency for the subset of events with $B$ rest frame
photon energy 
$E_\gamma > 1.5$ GeV ($M_{\Lambda \bar p} < 3.5$ GeV/$c^2$), and with
$E_\gamma > 2.0$ GeV ($M_{\Lambda \bar p} < 2.6$ GeV/$c^2$).  For 
$B^- \rightarrow \Lambda \bar p \gamma$, for events
with $E_\gamma > 1.5$ GeV we find efficiencies of 11.6\% (for $P/M$)
and 10.5\% (for $P^3/M$); for events with $E_\gamma > 2.0$ GeV we
find an efficiency of 12.4\% in both cases.  For the
$E_\gamma > 1.5$ GeV case, we conservatively use the smaller
efficiency.  For $B^- \rightarrow \Sigma^0 \bar p \gamma$, for events
with $E_\gamma > 1.5$ GeV we find efficiencies of 9.4\% (for $P/M$)
and 8.2\% (for $P^3/M$); for events with $E_\gamma > 2.0$ GeV we
find an efficiency of 10.6\% in both cases.  

There are also systematic errors in the efficiency from
uncertainty in the simulation of the detector performance
(track-finding, photon-finding, vertex-finding, resolutions) and an
uncertainty in
the modeling of the other $B$. We estimate these at $\pm 8.2$\%.

We obtain a conservative 90\% confidence level upper limit
on the branching fraction by using unpolarized $\Lambda$'s, using
the $P^3/M$ option for the $\Lambda \bar p$ ($\Sigma^0 \bar p$)
mass distribution, and
then increasing the limit so obtained by 1.28 times the quadratic
sum of the two remaining systematic errors, $\pm 2$\% from number of
$B$'s and $\pm 8.2$\% from detector simulation and modeling of the other
$B$.

While our specific goal in the first search was $B^- \rightarrow \Lambda
\overline p \gamma$, we also have sensitivity to the decay $B^-
\rightarrow \Sigma^0 \overline p \gamma$, $\Sigma^0 \rightarrow
\Lambda \gamma$ in that analysis.  Our efficiency for the latter decay
is 0.3 times that of the former.  Similarly, while our specific goal in
the second search was $B^- \rightarrow \Sigma^0 \bar p \gamma$, we also
have sensitivity to $B^- \rightarrow \Lambda \bar p \gamma$, 0.4 that
for $B^- \rightarrow \Sigma^0 \bar p \gamma$.
Hence, our primary results can be written

\begin{widetext}
$$[{\cal B} (B^- \rightarrow \Lambda \overline p \gamma) + 0.3
{\cal B} (B^- \rightarrow \Sigma^0 \overline p \gamma)]
_{E_\gamma > 1.5 \rm{~GeV}}  < 3.9 \times 10^{-6}\ ,$$

$$[{\cal B} (B^- \rightarrow \Lambda \overline p \gamma) + 0.3
{\cal B} (B^- \rightarrow \Sigma^0 \overline p \gamma)]
_{E_\gamma > 2.0 \rm{~GeV}}  < 3.3 \times 10^{-6}\ ,$$

$$[{\cal B} (B^- \rightarrow \Sigma^0 \overline p \gamma) + 0.4
{\cal B} (B^- \rightarrow \Lambda \overline p \gamma)]
_{E_\gamma > 1.5 \rm{~GeV}}  < 7.9 \times 10^{-6}\ ,$$

$$[{\cal B} (B^- \rightarrow \Sigma^0 \overline p \gamma) + 0.4
{\cal B} (B^- \rightarrow \Lambda \overline p \gamma)]
_{E_\gamma > 2.0 \rm{~GeV}}  < 6.4 \times 10^{-6}\ .$$
\end{widetext}

From these upper limits, we would like to obtain an upper
limit on the branching fraction for $b \rightarrow s \gamma$
leading to baryons.  Our first step in this direction uses isospin
considerations.  The parton-level final states, $s \bar u$ and
$s \bar d$, form an isospin doublet, and the hadronization process
should conserve isospin.  This gives
${\cal B}(B^- \rightarrow \Lambda \bar p \gamma) = 
{\cal B}(\bar B^0 \rightarrow \Lambda \bar n \gamma)$, and
${\cal B}(B^- \rightarrow \Sigma^0 \bar p \gamma) = 
{\cal B}(\bar B^0 \rightarrow \Sigma^0 \bar n \gamma) = 
1/2{\cal B}(B^- \rightarrow \Sigma^- \bar n \gamma) = 
1/2{\cal B}(\bar B^0 \rightarrow \Sigma^+ \bar p \gamma)$.  Thus
${\cal B}(B \rightarrow \Sigma \bar N \gamma) = 
3{\cal B}(B^- \rightarrow \Sigma^0 \bar p \gamma)$, and ${\cal B}(B
\rightarrow (\Lambda\ or\ \Sigma) \bar N \gamma) = 
3{\cal B}(B^- \rightarrow \Sigma^0 \bar p \gamma) + {\cal B}(B^-
\rightarrow \Lambda \overline{p} \gamma)$.  Multiplying
the last upper limit given above by 3, we have
$[{\cal B}(B \rightarrow \Sigma \bar N \gamma) + 1.2{\cal B}(B
\rightarrow \Lambda \overline{N}\gamma)]_{E_\gamma>2.0 \rm{~GeV}} < 1.9 \times
10^{-5}$, which we use as our limit on ${\cal B}(B \rightarrow (\Lambda\
or\ \Sigma)\overline{N} \gamma)$.

The above branching fraction limit is for both baryon and antibaryon
in the lowest lying baryon SU(3) octet.  We must also consider decays with
one of the baryons in the decuplet (i.e.,
$B \rightarrow \Sigma \bar \Delta \gamma$ and
$B \rightarrow \Sigma(1385) \bar N \gamma$), decays involving higher-mass
octet and decuplet members, and non-resonant decays such as 
$B \rightarrow (\Lambda\ or\ \Sigma)(\bar N\ or\ \bar \Delta) \pi
\gamma$.  The requirement that $E_\gamma(B\rm {~rest ~frame})$ be greater
than 2.0 GeV translates into an upper limit on the mass of the
baryon-antibaryon system of 2.60 GeV/$c^2$.  The various mass thresholds are:
$\Lambda \bar N$, 2.05 GeV/$c^2$; $\Sigma \bar N$, 2.13 GeV/$c^2$; $\Sigma \bar \Delta$,
2.43 GeV/$c^2$; $\Sigma(1385) \bar N$, 2.32 GeV/$c^2$.  Thus, phase space will suppress
the octet-decuplet rates relative to the octet-octet rates.  Combining
this with  the falling parton-level hadronic mass distribution given by the
spectator model \cite{Ali-Greub}, or the calculation of Kagan and
Neubert \cite{Anatomy}, we estimate a suppression of a factor of $\sim$4.
This is partially compensated by the factor of 2 more spin states
available in the octet-decuplet combination.  A plausible assumption
is that the octet-decuplet contribution would be $\sim$1/2 that of the
octet-octet contribution.  Octet-decuplet pairs with an excited member
are above the 2.60 GeV/$c^2$ cutoff imposed by the 2.0 GeV photon energy
requirement, and octet-octet pairs with excited members have thresholds
very close to the cutoff.  Non-resonant
$(\Lambda\ or\ \Sigma)(\bar N\ or\ \bar \Delta) \pi$ states will have
thresholds below the cutoff, but will be phase-space-suppressed
relative to $(\Lambda\ or\ \Sigma) \bar N$.  From all this, we take
as our working assumption
${\cal B}(b \rightarrow s \gamma,\ with\ baryons)_{E_{\gamma} > 2.0 \rm{~GeV}}
= 2{\cal B}(B \rightarrow (\Lambda\ or\ \Sigma) \bar N \gamma)
_{E_{\gamma} > 2.0 \rm{~GeV}}$, and hence
${\cal B}(b \rightarrow s \gamma,\ with\ baryons)_{E_{\gamma} > 2.0 \rm{~GeV}}
< 3.8 \times 10^{-5}$.

CLEO's recent study \cite{new} of $b \rightarrow s \gamma$ reported
a branching fraction for $E_\gamma > 2.0$ GeV, corrected for
the $b \rightarrow d \gamma$ contribution, of
$(2.94 \pm0.39 \pm0.25)\times 10^{-4}$.  Our upper limit on the
branching fraction for $b \rightarrow s \gamma$ leading to baryons,
with $E_\gamma > 2.0$ GeV, $3.8 \times 10^{-5}$, is 13\% of that
number.  The recent study \cite{new} had an efficiency for detecting
$B \rightarrow baryons\ \gamma$ that was 1/2 of that for modes not
involving baryons.  That implies an upper limit on the correction
needed for the branching fraction reported there of 6.5\%, less
than half the combined reported statistical ($\pm 13\%$) and systematic
($\pm 8\%$) errors.

CLEO's recent study \cite{new} of $b \rightarrow s \gamma$ also
reported information on the photon energy spectrum:  an average
energy $\langle E_\gamma \rangle = (2.346 \pm 0.032 \pm 0.011)$ GeV,
and a variance
$\langle (E_\gamma - \langle E_\gamma \rangle )^2 \rangle
= (0.0226 \pm 0.0066 \pm 0.0020)$ ${\rm GeV}^2$. Both averages were
taken {\it only for photons above 2.0 GeV}.  The average energy
of $B$ rest frame photons from events with baryons (averaging
only for photons above 2.0 GeV) is $\sim$2.1 GeV, 250 MeV lower
than the published mean.  The upper limit on the correction
to the first moment is thus 6.5\% of 250 MeV, i.e. 16 MeV
(compared with the published statistical and systematic
errors of 32 MeV and 11 MeV, respectively).  The limit
on the correction to the variance is 0.0025 ${\rm GeV}^2$, which
is 36\% of the combined quoted statistical and systematic errors
on the variance.

In conclusion, we have conducted searches for the exclusive
radiative penguin decays $B^- \rightarrow \Lambda \overline p \gamma$,
and $B^- \rightarrow \Sigma^0 \overline p \gamma$, 
found no evidence for either, and placed upper limits on them of

\begin{widetext}
$$[{\cal B} (B^- \rightarrow \Lambda \overline p \gamma) + 0.3
{\cal B} (B^- \rightarrow \Sigma^0 \overline p \gamma)]
_{E_\gamma > 1.5 \rm{~GeV}}  < 3.9 \times 10^{-6}\ ,$$

$$[{\cal B} (B^- \rightarrow \Lambda \overline p \gamma) + 0.3
{\cal B} (B^- \rightarrow \Sigma^0 \overline p \gamma)]
_{E_\gamma > 2.0 \rm{~GeV}}  < 3.3 \times 10^{-6}\ ,$$

$$[{\cal B} (B^- \rightarrow \Sigma^0 \overline p \gamma) + 0.4
{\cal B} (B^- \rightarrow \Lambda \overline p \gamma)]
_{E_\gamma > 1.5 \rm{~GeV}}  < 7.9 \times 10^{-6}\ ,$$

$$[{\cal B} (B^- \rightarrow \Sigma^0 \overline p \gamma) + 0.4
{\cal B} (B^- \rightarrow \Lambda \overline p \gamma)]
_{E_\gamma > 2.0 \rm{~GeV}}  < 6.4 \times 10^{-6}\ .$$
\end{widetext}

\noindent With plausible assumptions, this leads to
the conclusion that $b \rightarrow s \gamma$ decays with baryons
in the final state and $E_\gamma > 2.0$ GeV
constitute at most 13\% of all $b \rightarrow
s \gamma$ decays with $E_\gamma > 2.0$ GeV. 
With that limit, the upper limit on corrections to our recent
measurement \cite{new} of the $b \rightarrow s \gamma$ branching
fraction, the mean energy of the photon, and the variance
in the photon energy, are less than half of the combined quoted
statistical and systematic errors.

We gratefully acknowledge the effort of the CESR staff 
in providing us with
excellent luminosity and running conditions.
This work was supported by 
the National Science Foundation,
the U.S. Department of Energy,
the Research Corporation,
and the 
Texas Advanced Research Program.


\begin{thebibliography}{99}


\bibitem{Hurth}  
T.~Hurth,
hep-ph/0212304.

\bibitem{Anatomy}  
A.~L.~Kagan and M.~Neubert, 
Eur. Phys. J. C {\bf 7}, 5 (1999).

\bibitem{old} 
M.S. Alam {\it et al.} (CLEO), 
Phys. Rev. Lett.
    {\bf 74}, 2885 (1995).

\bibitem{Belle} 
K. Abe {\it et al.} (Belle), 
    Phys. Lett. B{\bf 511}, 151 (2001)
    [hep-ex/0103042].

\bibitem{new}  
S. Chen {\it et al.} (CLEO), 
Phys. Rev. Lett. {\bf 87}, 251807 (2001).

\bibitem{Ali-Greub}  
    A. Ali and C. Greub, Phys. Lett. {\bf B259}, 182 (1991); and private
    communications.

\bibitem{CLEO-detector} 
Y.~Kubota {\it et al.} (CLEO),  Nucl. Instrum. Methods
Phys. Res., Sect. A {\bf 320}, 66 (1992);
T.~Hill, Nucl. Instrum. Methods Phys. Res., Sect. A {\bf 418}, 32
(1998).

\bibitem{FW}  
G.~C.~Fox and S.~Wolfram,
Phys.\ Rev.\ Lett.\  {\bf 41}, 1581 (1978).

\bibitem{PDGlim}  
R.~M.~Barnett {\it et al.}  (Particle Data Group),
Phys.\ Rev.\ D {\bf 54}, 1 (1996).



%
%
%
%
%
%
%
%
%
\end{thebibliography}
\end{document}